\documentclass[twocolumn,superscriptaddress,letter]{revtex4}%
\usepackage{amssymb}
\usepackage{color}
\usepackage{graphicx}
\usepackage{dcolumn}
\usepackage{bm}
\usepackage[header,title,page,titletoc]{appendix}
\usepackage{amsmath}
\usepackage{amsfonts}%
\providecommand{\U}[1]{\protect \rule{.1in}{.1in}}

\begin{document}
\title{Particle-Hole Symmetry Protected Zero Modes on Vacancies in the Topological
Insulators and Topological Superconductors on the Honeycomb Lattice}
\author{Jing He}
\affiliation{Department of Physics, Beijing Normal University, Beijing, 100875, P. R. China}
\author{Ying-Xue Zhu}
\affiliation{Department of Physics, Beijing Normal University, Beijing, 100875, P. R. China}
\author{Ya-Jie Wu}
\affiliation{Department of Physics, Beijing Normal University, Beijing, 100875, P. R. China}
\author{Lan-Feng Liu }
\affiliation{Department of Physics, Beijing Normal University, Beijing, 100875, P. R. China}
\author{Ying Liang}
\affiliation{Department of Physics, Beijing Normal University, Beijing, 100875, P. R. China}
\author{Su-Peng Kou}
\thanks{Corresponding author}
\email{spkou@bnu.edu.cn}
\affiliation{Department of Physics, Beijing Normal University, Beijing, 100875, P. R. China}

\date{\today}

\begin{abstract}
In this paper we study the quantum properties of the lattice vacancy in
topological band insulators (TBIs) and topological superconductors (TSCs) on
honeycomb lattice with particle-hole symmetry. Each vacancy has one zero mode
for the Haldane model and two zero modes for the Kane-Mele model. In addition,
in TSCs on honeycomb lattice with particle-hole symmetry, we found the
existence of the Majorana zero modes around the vacancies. These zero energy
modes are protected by particle-hole symmetry of these topological sates.

\end{abstract}
\maketitle

Topological band insulators (TBIs) represent a class of novel states of matter
characterized by a special band structure : they are the materials that behave
like insulators in their interior or bulk while permitting the metallic
boundaries\cite{2,qhe,k,qi1}. In two dimensions, there are two types of TBIs,
the TBIs with time-reversal symmetry (TRS) and those without TRS. For the TBIs
without TRS\cite{haldane}, the Thouless-Kohmoto-Nightingale-Nijs (TKNN) number
can be identified to be the topological invariant\cite{thou}; While for the
TBIs with TRS (quantum spin Hall states in two dimension\cite{kane,zhang}),
people use the $Z_{2}$ topological invariant to label their properties
\cite{kane}. On the other hand, recently, people found that superconducting
(SC) states with the same local order parameter may have different topological
properties, which leading to the concept of "topological superconductivity
(TSC)"\cite{vol,read}.

The non-trivial topology of the bulk can also be exposed by introducing
topological defects, $\pi$-flux, dislocations and vortices, to name a few.
Moreover, the interplay of defect topology and topology of the original states
result in even richer phases. Previous works have been done on the effect of
$\pi$-flux and dislocations on the Haldane model which indeed trap zero energy
bound states\cite{franz,ju}. Moreover, according to the index
theorem\cite{index,index1,index2}, a vortex in topological superconductors is
predicted to harness a Majorana fermion zero mode inside the vortex
core\cite{vol}. In turn, such vortices were found to be non-Abelian anyons
which are the holy grail of topological quantum computation\cite{read}.

\begin{figure}[ptbh]
\includegraphics[width = 7.0cm]{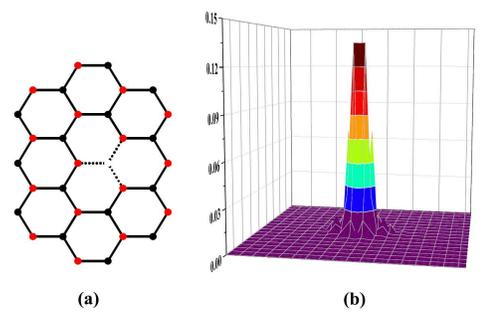}
\caption{(Color online) (a) The lattice defect of the honeycomb lattice -
vacancy. (b) The particle density of the zero mode around the vacancy of the
Haldane model with particle-hole symmetry ($\varepsilon=0$). }%
\label{1}%
\end{figure}

In this paper, we will focus on the effects of different type of
defects--vacancies, in two dimensional topological states with particle-hole
symmetry on the honeycomb lattice. FIG.1.a is the illustration of a vacancy on
the honeycomb lattice. It is known in graphene, zero modes of
electrons\textbf{ }were found around the vacancies\cite{0,1}. Here we also
found that the lattice defects of the TBIs on honeycomb lattice have
nontrivial quantum properties : for the Haldane model each vacancy has one
zero mode and for the Kane-Mele model each vacancy has two zero modes. These
zero energy modes in TBIs are protected by particle-hole symmetry and the
finite energy gap of the electrons. For the graphene system without energy
gap, the zero modes (localized states) of electrons around the vacancies are
fragile. By adding the perturbations that break the particle-hole symmetry
(for example, the staggered potential), these zero modes (localized states)
disappear. On the other hand, our results are much different from those in the
Ref.\cite{3}, in which the effect of lattice vacancies in graphene with
intrinsic spin-orbit interaction is studied in a continuum model. And {these
zero modes around a vacancy} cannot be simply regarded as the remnant of the
gapless edge states in a continuum effective model as people have done in
Ref.\cite{shen}. In addition, in TSCs on honeycomb lattice with particle-hole
symmetry, we found that there also exists (non-topological) Majorana zero
modes around the vacancies.

\textit{Vacancies for the Haldane model}: We start from the spinless Haldane
model, of which the Hamiltonian is defined as%
\begin{align}
H_{\mathrm{H}}  &  =-t\sum \limits_{\left \langle {i,j}\right \rangle }\left(
\hat{c}_{i}^{\dagger}\hat{c}_{j}+h.c.\right)  -t^{\prime}\sum
\limits_{\left \langle \left \langle {i,j}\right \rangle \right \rangle }%
e^{i\phi_{ij}}\hat{c}_{i}^{\dagger}\hat{c}_{j}\nonumber \\
&  +\varepsilon \sum \limits_{i\in{A}}\hat{c}_{i}^{\dagger}\hat{c}%
_{i}-\varepsilon \sum \limits_{i\in{B}}\hat{c}_{i}^{\dagger}\hat{c}_{i}%
\end{align}
where $t$ and $t^{\prime}$ are the nearest-neighbor hopping and the
next-nearest-neighbor hopping, respectively. $e^{i\phi_{ij}}$ is a complex
phase of the next-nearest-neighbor hopping, and we set the direction of the
positive phase to be clockwise $\left(  \left \vert \phi_{ij}\right \vert
=\frac{\pi}{2}\right)  $. $\varepsilon$\ is\ the\ coefficient of
the\ on-site\ staggered\ energy.$\ $Using the Fourier transformations, we can
get the spectrum of free fermions as $E_{k}=\pm \sqrt{(\xi_{k}^{\prime
}-\varepsilon)^{2}+\left \vert \xi_{k}\right \vert ^{2}}$ where $\left \vert
\xi_{k}\right \vert =t\sqrt{3+2\cos{(\sqrt{3}k_{y})}+4\cos{(3k_{x}/2)}%
\cos{(\sqrt{3}k_{y}/2)}}$ and $\xi_{k}^{\prime}=2t^{\prime}\left[
-2\cos({3k_{x}/2})\sin({\sqrt{3}k_{y}/2)+\sin(\sqrt{3}k_{y})}\right]  .$ For
this free fermionic system, there are two phases: the TBI state with TKNN
number $C=\pm1$ and the normal band insulator (NI) state. The phase boundary
is $2\varepsilon=6\sqrt{3}t^{\prime}.$

Firstly we study the particle-hole symmetry of the Haldane model. After doing
the particle-hole transformation, $\hat{c}_{i\in A}^{\dagger}\leftrightarrow
-\hat{c}_{i\in A}$ and $\hat{c}_{i\in B}^{\dagger}\leftrightarrow \hat{c}_{i\in
B}$ together with a complex (or charge) conjugate transformation\cite{yang},
the hopping terms are invariant but the staggered potential term is changed.
So for the case of $\varepsilon=0,$ the Hamiltonian have the particle-hole
symmetry as $H_{\mathrm{H}}=P^{\dagger}H_{\mathrm{H}}P$ where $P$ is the
particle-hole transformation operator for spinless fermion on the honeycomb
lattice. Now we have \emph{the} \emph{spectrum symmetry} : each energy level
of the electrons with positive energy $E$ must be paired with an energy level
of the electrons with negative energy $-E$. While for the case of
$\varepsilon \neq0,$ the particle-hole symmetry is broken, $H_{\mathrm{H}}\neq
P^{\dagger}H_{\mathrm{H}}P$. However, now, we still have the spectrum symmetry
as $E\Leftrightarrow-E$. This is because after doing the particle-hole
symmetry, the staggered potential changes sign. Due to the translation
symmetry, we may do a sublattice transformation as $A\longleftrightarrow B$.
Then the total Hamiltonian is invariant.

We study the effect of a vacancy on the electronic states. It is obvious that
such a defect that breaks translation symmetry is a local distortion of the
system and has no topological properties. We can consider a vacancy as a
"hole" in the system by removing a site. Due to the non-zero TKNN number,
there may exist topologically protected edge states on the boundary of this
"hole". It is known that at low energy the dispersion of edge states has a
form as $E(k)\sim v_{F}k$ where $v_{F}$ is Fermi velocity of edge states. When
the size of the hole shrinks, the energy levels of the edge states become
discrete and eventually the edge states on the boundary around the "hole" turn
into localized states around the vacancy. Because the Haldane model is based
on a bipartite lattice, when we remove a site to create a vacancy, there will
exist an unpaired electronic state. Due to the spectrum symmetry
($E\Leftrightarrow-E$), the corresponding localized unpaired electronic state
must have exact zero energy. Such zero mode is protected by the particle-hole
symmetry of the TBIs.

We calculate the electronic states for the Haldane model with a vacancy for
the particle-hole symmetry case ($\varepsilon=0$) numerically. On a
$36\times36$ lattice, we found a zero mode near the vacancy in the Haldane
model. In FIG.1.b, we plot the particle density of the zero modes. The
particle density is localized around the defect center within a length-scale
$\sim(\Delta_{f})^{-1}$ where $\Delta_{f}$ is the fermion's energy gap.
FIG.2.a shows the energy levels via $t^{\prime}$ for the Haldane model with a
vacancy. Now the total number of the electronic states for this case is odd.
FIG.2.b shows the density-of-state (DOS) of the electronic states, of which a
$\delta$ function is around $\omega=0$ due to the zero mode of the vacancy.

\begin{figure}[ptb]
\includegraphics[width=0.75\textwidth]{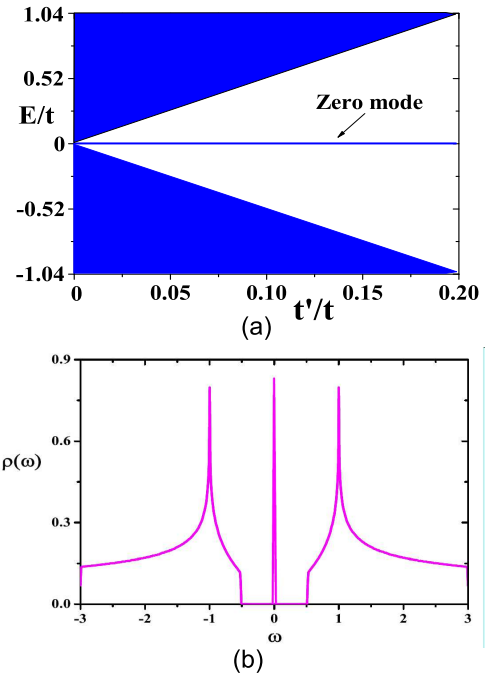}\caption{(Color online) (a)
The energy levels withthe spectrum symmetry ($E\Leftrightarrow-E$): the
localized state has zero energy shown by the blue line and the continuum.
spectrum is shown by the blue region (b) The scheme of DOS of electronic
states with a vacancy. In non-interacting limit, the contribution of DOS from
the vacancy is a $\delta$ function at $\omega=0$. Here we set $t^{\prime
}=0.1.$}%
\end{figure}

Besides, we calculate the induced quantum number on a vacancy. From the
numerical results, we have found a zero mode on the vacancy for the Haldane
model. So a vacancy possesses two localized states which are denoted by
$|+\rangle$ and $|-\rangle$ that are the occupied and unoccupied state of
fermions, respectively. Around a vacancy, the fermionic operators are expanded
as $\hat{\psi}(r,t)=\sum_{k\neq0}c_{k}e^{-iE_{k}t}\Psi_{k}(r)+\sum_{k\neq
0}d_{k}^{\dagger}e^{iE_{k}t}\Psi_{k}^{\dagger}(r)+c_{0}\Psi_{0}(r),$ where
$c_{k}$ and $d_{k}^{\dagger}$ are the operators of $k\neq0$ modes that are
irrelevant to low energy physics. $\Psi_{0}(r)\ $is the wave-function of the
unpaired zero mode. $c_{0}$ is the annihilation operator of the zero mode. We
define the induced fermion number operators as $\hat{N}_{F}\equiv \int
:\hat{\psi}^{\dagger}\hat{\psi}:d^{2}\mathbf{r}=c_{0}^{\dagger}c_{0}%
+\sum \limits_{k\neq0}(c_{k}^{\dagger}c_{k}-\hat{d}_{k}^{\dagger}\hat{d}%
_{k})-\frac{1}{2}$ where $:\hat{\psi}^{\dagger}\hat{\psi}:$ means normal
product of $\hat{\psi}^{\dagger}\hat{\psi}$. So we obtain the eigenvalues of
induced fermion number operator as $\hat{N}_{F}|\pm \rangle=\pm{\frac{1}{2}%
}|\pm \rangle.$ The occupation (or unoccupation) of this zero mode lead to
$N_{F}=\frac{e}{2}$ (or $N_{F}=-\frac{e}{2}$) fractional electronic charge.
Although there is no topological background, an induced fractional charge is
trapped around each vacancy in the Haldane model.

In addition, we found \emph{the parity effect} for the Haldane model with
multi-vacancy: for odd number vacancies (for example, 3 vacancies, 5
vacancies...), there exist zero energy modes; while for even number
multi-vacancy (for example, 2 vacancies, 4 vacancies...), the localized states
will have finite energy. It is also the particle-hole symmetry that enforces a
zero energy level together with the pairs of energy levels with finite energy
for the case of odd number vacancies. For example, for two separated
vacancies, there exist two localized states (quasi-zero-modes) and the
localized states slightly split due to the quantum tunneling effect between
them. It is obvious that the energy splitting decays exponentially as the
distance between two vacancies increases.

\begin{figure}[ptb]
\includegraphics[width=0.5\textwidth]{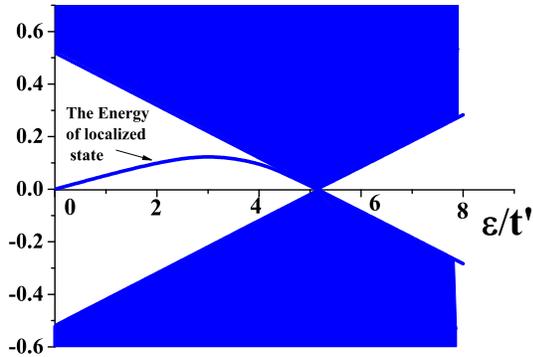}\caption{The energy of the
localized states around the vacancies via the staggered potential
$\varepsilon$. For $\varepsilon \neq0$, there is no spectrum symmetry
($E\nLeftrightarrow-E$) and the localized state has a finite energy shown by
the blue line (the continuum spectrum is shown by the blue region). For
$\varepsilon>3\sqrt{3}t^{\prime},$ the ground state is NI phase, there is no
localized state around the vacancy.}%
\end{figure}

We have found the particle-hole symmetry protected zero mode around vacancies
in the TBI phase of the Haldane model. Then we add the staggered on-site
potential which breaks particle-hole symmetry. {When the particle-hole
symmetry is broken, the localized state will shift from zero to a finite
value.} We plot the bound state energy via $t^{\prime}$ for the case of
$\varepsilon=0.1t$ in FIG.3. For $\varepsilon<\varepsilon_{c}=3\sqrt
{3}t^{\prime},$ in TBI phase, there exist a localized state around the
vacancy. For $\varepsilon>\varepsilon_{c}=3\sqrt{3}t^{\prime},$ the ground
state is NI phase with trivial quantum properties, then the localized states disappear.

Finally, by varying local disordered potential on a given site $i_{0}$
gradually, the wave-function evolves from the extended state to the localized
state. The Hamiltonian model in Eq.(1) turns into
\begin{align}
H_{\mathrm{H-V}}  &  =-t\sum \limits_{\left \langle {i{\neq}i_{0},j\neq}%
i_{0}\right \rangle }\left(  \hat{c}_{i}^{\dagger}\hat{c}_{j}+h.c.\right)
-t^{\prime}\sum \limits_{\left \langle \left \langle {i{\neq}i_{0},j\neq}%
i_{0}\right \rangle \right \rangle }e^{i\phi_{ij}}\hat{c}_{i}^{\dagger}\hat
{c}_{j}\nonumber \\
&  -\alpha t\sum \limits_{\left \langle i_{0}{,j}\right \rangle }\left(  \hat
{c}_{i}^{\dagger}\hat{c}_{j}+h.c.\right)  -\alpha t^{\prime}\sum
\limits_{\left \langle \left \langle i_{0}{,j}\right \rangle \right \rangle
}e^{i\phi_{ij}}\hat{c}_{i}^{\dagger}\hat{c}_{j}\nonumber \\
&  +V_{0}\hat{c}_{i_{0}\in{A}}^{\dagger}\hat{c}_{i_{0}\in{A}}.
\end{align}
where $V_{0}=(1-\alpha)t/\alpha.$ $\alpha$ is a parameter to tune the local
disordered potential. In the limit of $\alpha \rightarrow0,$ the on-site
potential turns into infinite and the hopping parameters turn to zero around
the site $i_{0}$. On the other hand, in the limit of $\alpha \rightarrow1,$ the
system has translation invariance, of which the Hamiltonian reduces into
Eq.(1) of $\varepsilon=0$. By tuning $\alpha,$ we don't find any quantum phase
transition. One extended state will turn into the localized state when there
exists arbitrary small disordered potential with $\alpha \neq0$. See FIG.4.

\begin{figure}[ptb]
\includegraphics[width=0.5\textwidth]{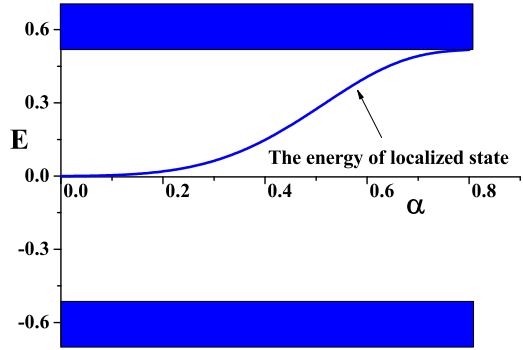}\caption{The energy of the
localized state around a vacancy via the on-site disordered parameter $\alpha
$. For $\alpha \neq0$, there is no spectrum symmetry ($E\nLeftrightarrow-E$)
and the localized state has a finite energy shown by the blue line (the
continuum spectrum is shown by the blue region).}%
\end{figure}

On the other hand, we studied the properties of the topological defects (the
$\pi$-flux on a plaquette of the honeycomb lattice) in the Haldane model and
also found a zero mode around it in the TBI phase. The zero modes around the
$\pi$-flux are protected by the topological invariable. In TBI phase, the
localized state has exact zero energy which is robust against arbitrary
perturbation. There also exists an induced fractional charge $N_{F}=\frac
{e}{2}$ around the $\pi$-flux. With the induced fractional charge, each $\pi
$-flux carries a statistical angle $\pi/4$.

\textit{Vacancies for the Kane-Mele model}: The Hamiltonian of the Kane-Mele
model is similar to that of the Haldane model as%
\begin{align}
H_{\mathrm{KM}}  &  =-t\sum \limits_{\left \langle {i,j}\right \rangle ,\sigma
}\left(  \hat{c}_{i,\sigma}^{\dagger}\hat{c}_{j,\sigma}+h.c.\right)
-t^{\prime}\sum \limits_{\left \langle \left \langle {i,j}\right \rangle
\right \rangle }e^{i\phi_{ij}}\hat{c}_{i}^{\dagger}\sigma_{z}\hat{c}%
_{j}\nonumber \\
&  +\varepsilon \sum \limits_{i\in{A,\sigma}}\hat{c}_{i,\sigma}^{\dagger}\hat
{c}_{i,\sigma}-\varepsilon \sum \limits_{i\in{B,\sigma}}\hat{c}_{i,\sigma
}^{\dagger}\hat{c}_{i,\sigma}%
\end{align}
where $\sigma_{z}$ is the Pauli matrix and $\sigma$ are the spin-indices
representing spin-up $(\sigma=\uparrow)$ and spin-down $(\sigma=\downarrow)$
for electrons. Using similar calculations, we found that there exist two zero
modes around a vacancy for the Kane-Mele model with $\varepsilon=0$ and there
also exists the parity effect for multi-vacancy case. These zero modes are
also protected by the particle-hole symmetry. When there exist particle-hole
breaking terms, the zero modes turn into the localized modes with finite energy.

Then we calculate the induced quantum number on a vacancy for the Kane-Mele
model. Using similar approach, we found that for the TBI phase, the two zero
modes around a vacancy correspond to four degenerate energy levels denoted by
$\mid \uparrow_{+}\rangle \otimes \mid \downarrow_{+}\rangle,$ $\mid \uparrow
_{-}\rangle \otimes \mid \downarrow_{-}\rangle,$ $\mid \uparrow_{-}\rangle
\otimes \mid \downarrow_{+}\rangle,$ $\mid \uparrow_{+}\rangle \otimes
\mid \downarrow_{-}\rangle$\cite{kou1}. Here $\mid \sigma_{+}\rangle$ and
$\mid \sigma_{-}\rangle$ are the occupied and unoccupied state, respectively.
For the half filling case, the localized states around the vacancy are
$\mid \uparrow_{-}\rangle \otimes \mid \downarrow_{+}\rangle,$ $\mid \uparrow
_{+}\rangle \otimes \mid \downarrow_{-}\rangle,$ of which there exists
a\ spin-$\frac{1}{2}$\ moment as
\begin{align}
\hat{S}^{z}  &  \mid \uparrow_{-}\rangle \otimes \mid \downarrow_{+}\rangle
=\frac{1}{2}\mid \uparrow_{-}\rangle \otimes \mid \downarrow_{+}\rangle,\text{
}\nonumber \\
\hat{S}^{z}  &  \mid \uparrow_{+}\rangle \otimes \mid \downarrow_{-}\rangle
=-\frac{1}{2}\mid \uparrow_{+}\rangle \otimes \mid \downarrow_{-}\rangle.
\end{align}
In Ref.\cite{yao}, people proposed that the Kane-Mele model can be realized in
a Silicene based material with honeycomb lattice and strong spin-orbital
coupling interaction. Then in this system people can detect the local spin
moment by observing uniform spin susceptibility which obeys Curie-Weiss law as
$\chi_{\text{s}}\sim N_{\mathrm{ls}}(k_{B}T)^{-1}$ where $N_{\mathrm{ls}}$ is
the vacancy number.

Recently, people have studied the quantum properties of the $\pi$-flux in
Ref.\cite{qi,ran,kou1}. They found there exist two zero modes and induced
spin-$\frac{1}{2}$\ moment on the $\pi$-flux.

\textit{Vacancies for the TSCs on honeycomb lattice with particle-hole
symmetry}: In this part we will consider the effect of a vacancy in a
$\mathcal{C}=\pm1$ chiral TSCs on honeycomb lattice with particle-hole
symmetry. The effective model of the TSC is
\begin{equation}
H_{\mathrm{TSC}}=H_{\mathrm{H}}+\Delta_{\mathrm{induce}}\sum_{\left \langle
ij\right \rangle }\hat{c}_{i}\hat{c}_{j}+h.c.
\end{equation}
where $H_{\mathrm{H}}$ is the Hamiltonian of the Haldane model with
$\varepsilon=0$ and $\Delta_{\mathrm{induce}}$ is the induced SC order
parameter due to the proximity effect to an extended s-wave SC order. Now the
ground state is really a $\mathcal{C}=\pm1$ TSC, of which the topological
properties are similar to those of $p_{x}+\mathrm{i}p_{y}$ wave pairing
TSC.\textbf{ }Above effective model has particle-hole symmetry as,
$E\Leftrightarrow-E$.

We then studied the properties of a vacancy by solving the Bogolubov-de Gennes
(BdG) equations. On each vacancy, we found a single zero-mode which is
described by a real fermion field $\gamma^{\dagger}=\int d{r}[u_{0}%
\psi^{\dagger}+v_{0}\psi]$ ($\gamma^{\dagger}=\gamma$) obtained as a solution
of the BdG equations. When two vacancies fuse (taken to the nearby sites in
the honeycomb lattice), the result contains more than one quasi-particle due
to the Ising fusion rule \cite{mr}: $\sigma \times \sigma=I+\psi$ where $\sigma$
denotes a vacancy, $I$ the vacuum and $\psi$ the fermion. This fusion rule is
same to that of the non-Abelian anyons. However, although trapping a Majorana
zero mode, the vacancy has trivial quantum statistics and cannot be regarded
as a non-Abelian anyon. That is we found a symmetry protected Majorana zero
modes in two dimensions without topological defects. So we call it
\emph{non-topological Majorana mode}.

Furthermore, we studied the properties of the $\pi$-flux for this system and
also found a zero mode around of it in the TSC phase. So in the TSC state, the
localized state has exact zero energy which is also robust against arbitrary
perturbation. With trapping the Majorana zero modes, each $\pi$-flux obeys the
non-Abelian statistics.

\textit{Conclusion}: In the end we draw a conclusion. In general, for a TBI or
a TSC on a bipartite lattice with particle-hole symmetry, there must exist
zero energy modes around a vacancy and the parity effect for multi-vacancy. In
this paper we investigate the lattice vacancy by using TBIs and TSCs on
honeycomb lattice as examples to show this effect. For the TBIs and TSCs on a
$\pi$-flux lattice (another two dimensional bipartite lattice), the properties
of vacancies are similar to those on honeycomb lattice. As the remnant of the
gapless edge state around a "hole" with minimize size, the localized states
around the vacancies will have exact zero energy which are protected by the
particle-hole symmetry of these topological states. The perturbations breaking
particle-hole symmetry will shift the zero energy of the localized state to a
finite value. But the induced quantum numbers on the vacancies doesn't change
until the energy gap closes and the quantum transition phase occurs.

Finally we give a table to show the difference between the zero modes around a
topological defect (the $\pi$-flux) and the the zero modes around a vacancy
(non-topological defect) in difference topological states.

\begin{widetext}
\begin{table*}[t]%
\begin{tabular}
[c]{|c|cccc|}\hline & The Haldane model & The Kane-Mele model & The TSC
on honeycomb lattice\\
\hline
$\pi$-flux& Abelian anyon with statistical angle $\pi/4$ & spin moment & Non-Abelian anyon\\
vacancy& e/2 charge
& spin moment & Majorana fermion mode &\\ \hline
\end{tabular}
\caption{In different topological states, the $\pi$-flux (topological defect) and the vacancy (non-topological defect) have different properties.}
\end{table*}
\end{widetext}

\begin{acknowledgments}
This work is supported by National Basic Research
Program of China (973 Program) under the grant No. 2012CB921704, 2011CB921803, 2011cba00102 and NFSC Grant No. 11174035.
\end{acknowledgments}

\end{document}